\documentclass[aps,prl,twocolumn,amsmath,amssymb,floatfix]{revtex4}
\usepackage{tabularx}
\usepackage{bm}
\usepackage{euscript}
\usepackage{graphicx}
\usepackage{color}
\usepackage{amsfonts}
\usepackage{exscale}
\usepackage{amsbsy}

\begin{document}

\title{Nature of a topological quantum phase transition in a chiral spin liquid model}

\author{$^1$Suk Bum Chung, $^1$Hong Yao, $^1$Taylor L. Hughes, and $^2$Eun-Ah Kim}
\affiliation{$^1$ Department of Physics, Stanford University, Stanford, CA 94305\\
             $^2$ Department of Physics, Cornell University, Ithaca, NY 14853}
\date{\today}

\begin{abstract}
We study the finite temperature nature of a quantum phase transition between an
Abelian and a non-Abelian topological phase in an exactly solvable
model of a chiral spin liquid \cite{hongCSL}. By virtue of the exact solvability, this model can serve as a testbed for
developing better measures for describing topological quantum phase transitions.
 We characterize 
 this  phase transition in terms of the global flux and entanglement entropy, and
 discuss to what extent the existence of a topologically ordered ground state with non-Abelian excitations is revealed at finite temperature.
\end{abstract}

\maketitle 

\begin{figure}[bht] 
   \centering
   \includegraphics[width=.4\textwidth]{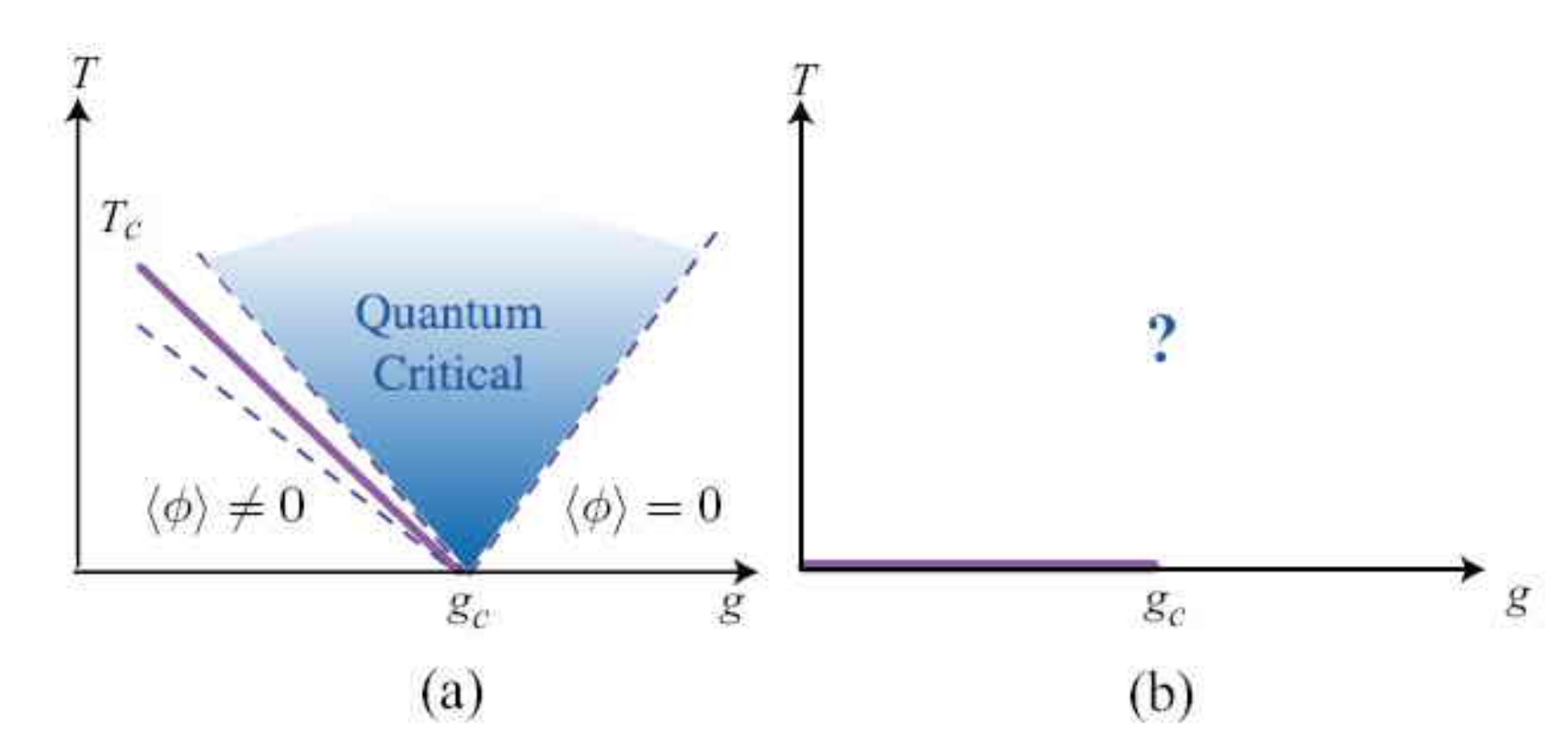}
   \caption{Phase diagrams in $g-T$ phase space. (a)A typical QPT phase diagram for conventional order with  the quantum fluctuations controlled by tuning parameter $g$. Here we sketched a case with the dynamical critical exponent $z=1$. (b)Topological QPT phase diagram. }
   \label{fig:PD}
\end{figure}

Characterizing and detecting topological order is one
of the central questions in the field of topological phases. The
challenge lies in that these new type of quantum ground states are
not associated with any local broken symmetry. Of broader interest
in the context of quantum phase transitions(QPT) is a question of
the nature of a quantum critical point when a system enters a
topologically ordered phase \cite{topoQPT}. In a conventional QPT, in which a
local order parameter starts to gain 
an expectation value at the quantum
critical point (QCP),  the nature of the QCP is of ultimate
significance. Even though it is effectively a point of measure
zero, it governs a much larger phase space often called the
``quantum critical region'' (See Fig.\ref{fig:PD}(a) where we
denote the expectation value of an order parameter by
$\langle\phi\rangle$). One can ask if, and to what extent, an
analogy holds for topological quantum phase transitions. For such
a question, we need  a formulation and understanding of measures
of topological order at finite temperature.

Since Wen and Niu \cite{wen-niu} coined the term ``topological
order" in association with the ground state degeneracy of
(Abelian) fractional quantum Hall (FQH) states on topologically
non-trivial surfaces,  such ground state degeneracy has been
widely used as an indicator of topological order including the
non-Abelian FQH states \cite{moore-read}. Further, the implications
of such degeneracy on fractionalization  has also been
discussed\cite{odd,frac_deg}. However, the  extension of this
indicator, which is 
defined at $T=0$ and not directly accessible 
experimentally, to a measure at finite temperature is an open
question.

More recently, the concept of ``topological entanglement entropy"
has been gaining interest as an indicator of topological
order \cite{kitaev-preskill,levin-wen} or a topological
QCP \cite{moore-fradkin}.  The corresponding quantity at finite
temperature has also been studied \cite{thermal_Claudio}. However,
one of the issues with topological entanglement entropy  is that
it does not always distinguish phases with obviously different
topological orders such as weak pairing (vortices follow
non-Abelian statistics) and strong pairing (vortices are Abelian)
$p+ip$ superconductors \cite{read-green}. This shows that the
information about the ground state is significantly condensed upon
mapping to a single entropic quantity.

We start with the observation that a chiral spin liquid (CSL)
model \cite{hongCSL} can  provide an ideal testbed for developing
a better understanding of a topological QPT by playing the role
that the transverse field Ising chain played in the study of
conventional QPTs. It has the virtue of being exactly solvable,
 and exhibiting a non-trivial QPT between non-Abelian($nA$) and
Abelian($A$) phases analogous to the weak pairing and strong
pairing limits of a $p+ip$ superconductor; the same physics can also be accomplished 
for the honeycomb model with three-spin interaction 
\cite{lee_soliton}. We take a twofold approach: First, we employ
the notion of an ``expectation value'' of a global flux operator
introduced by Nussinov and Ortiz \cite{fragility} as a finite
temperature extension of the concept of ground state degeneracy.
Second, we contrast this result to what can be learned from
entanglement entropy.

\begin{figure}
\centerline {
\includegraphics[width=.4\textwidth]{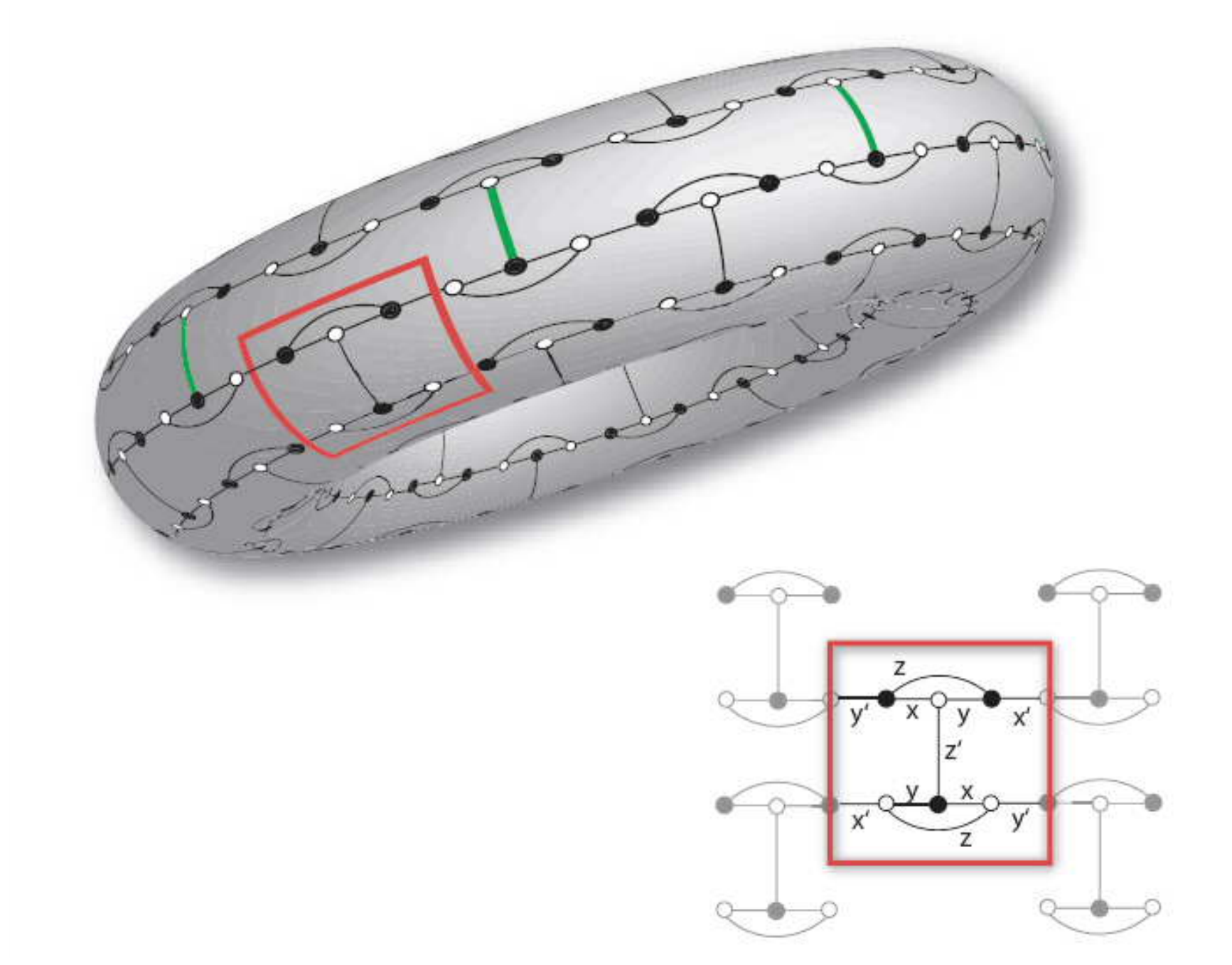}
}
\caption{A decorated brick wall lattice that is topologically equivalent to the star lattice of Ref.\cite{hongCSL},  on the surface of a torus. Green links denote $u_{ij}$ configurations contributing to a global flux threading. The inset defines  $\alpha$ links for  `triangles'  and $\alpha'$ links connecting `triangles' with $\alpha=x,y,z$. This labelling of links specifies components of spins interacting across the links in the original spin model.}
\label{FIG:TLattice}
\end{figure}

{\bf Model}-- The exactly solvable CSL model  on the star lattice
\cite{hongCSL} is a variant of a spin model with topological order
first introduced by Kitaev\cite{kitaev_rev} on the honeycomb
lattice. In this variation, ground states spontaneously break time
reversal symmetry and a QPT between $A$ and $nA$ phases is
accessible through the exact solution. For brevity we employ a
Majorana fermion representation of the model. We represent
spin-1/2 Pauli operators $\sigma^\alpha_i$ ($\alpha=x,y,z$)  of
the original spin model at each lattice site $i$\cite{hongCSL} by
four species of Majorana fermions $c_i$ and $d^\alpha_i$,
$\sigma^\alpha_i = i c_i d^\alpha_i$ under the constraint
\begin{equation}
D_i\equiv c_i d^x_i d^y_i d^z_i =1
\label{eq:constraint}
\end{equation} so that $\sigma^x\sigma^y\sigma^z=i$ as is expected of spin $1/2$ operators.
In terms of these Majorana fermions, the Hamiltonian is
\begin{eqnarray}
\mathcal{H}[\{\hat{U}_{ij}\}] = J\!\!\!\sum_{x,y,z{\rm-link}} \hat U_{ij}ic_i c_j + J'\!\!\!\sum_{x',y',z'{\rm-link}}\!\!\! \hat U_{ij}ic_i c_j,
\label{EQ:QuadHam}
\end{eqnarray}
where $\hat{U}_{ij}\equiv -id^\alpha_i d^\alpha_j$ is defined at
each $\alpha$ and $\alpha'$ bonds between sites $(i,j)$  (see
Fig.\ref{FIG:TLattice})
 and acts as a 
$\mathbb{Z}_2$ gauge field living on the $ij$ bond \footnote{Note that
the original representation in Ref. \onlinecite{hongCSL} introduces $\mathbb{Z}_2$ gauge
fields only on $z$, $z'$, and the `cut' links, which corresponds to a
particular gauge choice here.}. As
$\hat{U}_{ij}$ has no dynamics ($[\hat U_{ij},\mathcal{H}]=0$) it
can be replaced by a set of $\mathbb{Z}_2$ variables $u_{ij}=\pm
1$ reducing Eq. \eqref{EQ:QuadHam}
 to a quadratic Hamiltonian $\mathcal{H}[\{u_{ij}\}]$ parameterized by $\{u_{ij}\}$. 
For a loop $L$, the $Z_2$ flux is given by $\phi_L=\prod_{ij \in L} u_{ij}=\pm1$. 

Defining $g\equiv J'/J$, $\mathcal{H}[\{u_{ij}\}]$ can be
diagonalized as
\begin{equation}
\mathcal{H}[\{u_{ij}\}] = J\sum_{n, \vec{k}} \epsilon_{n,\vec{k}}[\{ u_{ij}\}; g]\, (b_{n,\vec{k}}^\dagger b_{n,\vec{k}} - 1/2),
\label{EQ:freeModes}
\end{equation}
by finding the complex fermion operators $b_{n, \vec{k}}$ that are
linear in $c_i$'s for momentum $\vec{k}$ and band index $n=1,2,3$
(there are six sites per unit cell in the Majorana fermion
Hamiltonian). This yields the entire spectrum
$\epsilon_{n,\vec{k}}[\{ u_{ij}\}; g]$.  
The ground states are uniform flux states with $\phi^0_L=-1$ for all 12-plaquettes and $\phi^0_L=1$ or $-1$ for all triangular plaquettes, spontaneously breaking time reversal symmetry. A vortex on a plaquette $L$, defined by $\phi_L = -\phi^0_L$, costs finite energy for all $g$. 
Moreover, the uniform flux ground state is degenerate on a torus
and  {\it the topological degeneracy changes} across the $nA$ to
$A$ QPT at $g_c=\sqrt{3}$\cite{hongCSL}. However, care is needed
for discerning physical states that satisfy the constraint Eq.
\eqref{eq:constraint} for each configuration of $\{u_{ij}\}$.

{\bf Topological degeneracy and the projection operator}-- A clue
towards an extension of topological degeneracy to finite
temperature lies in the $g$-dependent effects of the constraint
Eq. \eqref{eq:constraint}. The constraint defines the
\emph{physical} states  of the free fermion Hamiltonian
Eq. \eqref{EQ:freeModes} and is sensitive to $g$. 
Eq. \eqref{eq:constraint} can be implemented using
 a  projection operator $\hat{P} = \prod_i \frac{1}{2}(1+ D_i )$ since $D_i \hat{P} =\hat{P}$ \cite{kitaev_rev,hongCSL,Yao09}. Moreover, $\hat {P}$ commutes with the original
 Hamiltonian $\mathcal{H}[\{\hat{U}_{ij}\}]$.  
We can show that
whether a state survives projection only depends on the fermion parity, defined as $P_f=\prod_{ij\in x',y'z'\textrm{-links}} i c_{i} c_{j}$, and the parity of the number of vortex excitations \footnote{After gauge fixing, the $\hat{P}$ expansion of Ref. \cite{Yao09} can be factorized: $\hat{P} = [1 + \prod_i D_i]\hat{G} = [1+P_f \prod_{L\in \triangle,\triangledown} \phi_L] \hat{G}$, where $\hat{G} = [1 + \sum_j D_j + \sum_{i<j}D_i D_j+\cdots]$ (note $D_i^2=1$).} on triangle plaquettes. 

In the uniform flux sector, all physical states have even fermion parity $P_f=1$, which is 
particularly important for determining the topological degeneracy
of the ground states on a torus. Topological degeneracy comes from the identical free fermion spectra, in the thermodynamic limit, in the four possible topological sectors, distinguished by the choice of
the $\mathbb{Z}_2$ global flux:
\begin{eqnarray}
\Phi_\alpha\equiv \prod_{\langle ij\rangle \in \Gamma_\alpha}\! u_{ij}=\pm 1,
\label{eq:GF}
\end{eqnarray}
where $\alpha=x,y$ label two global cycles $\Gamma_x$ and
$\Gamma_y$. Now $(\Phi_x,\Phi_y)=(\pm1,\pm1)$ are four distinct
states  where $\Phi_x = -1$ and $\Phi_y=-1$ indicates $\pi$ flux
threaded through the distinct holes of the torus (see
Fig.\ref{FIG:TLattice}). 
Normally the fermion parity in the unprojected ground state wave function in all four topological sectors is even and independent of $g.$ Thus, they will survive the projection and give rise to four-fold topological degeneracy. This is indeed the case 
on the $A$ side
$(g>g_c)$. However, the fermion parity in the unprojected ground state wave function 
in the $(-1, -1)$ sector on
the $nA$ side $(g<g_c)$ is odd and consequently it does not survive projection \footnote{The exclusion of the $(-1,-1)$  state from the ground
state for $g<g_c$ is  tied to the $nA$ statistics of
the vortices. A global flux  threading $\Phi_\alpha$ is equivalent
to the procedure of (i)creating  a vortex pair (ii) transporting
one vortex around the loop $\Gamma_\alpha$(iii) annihilating the
pair\cite{odd}. The $(-1,-1)$  sector is equivalent to two vortex loops linked to each other, which cannot be undone in the $nA$ phase.}. Thus there is only a three-fold topological ground state degeneracy in the $nA$ phase. In summary, for $A$ phases with uniform flux, all the physical states consist of an even number of fermionic quasiparticle excitations above the ground states in all four topological sectors. For $nA$ phases with uniform flux, all physical states have an even number of fermion excitations in sectors $(1,1)$, $(1,-1)$, and $(-1,1)$, but an odd number of fermion excitations in the sector $(-1,-1).$ This has consequences not only for the topological ground state degeneracy, but also at finite temperature, as shown below.

{\bf Global flux expectation value}-- Motivated by the connection
between the change in the allowed physical spectrum at the
topological QPT and the global flux states, we consider the
`expectation value' of the global flux $\langle
\Phi_\alpha\rangle$ defined as \cite{fragility}
\begin{equation}
\langle \Phi_\alpha (T)\rangle\equiv\ \frac{1}{\mathcal{Z}}{\rm tr}\  \Phi_\alpha e^{-\mathcal{H}/T}
\label{eq:Phi}
\end{equation}
in a finite size system with $N$ sites. 
This ties the topological degeneracy to the spectrum and offers a
natural finite $T$ extension of topological degeneracy. If we
further restrict ourselves to uniform flux states (which is valid
at $T=0$ and is a good approximation in the vicinity of the QCP
where the fermion gap vanishes but vortex gap is finite), Eq.
\eqref{eq:Phi} can be recast as
\begin{equation}
\langle \Phi_\alpha \rangle =\frac{\sum_{\Phi_x,\Phi_y =\pm1}\Phi_\alpha\mathcal{Z}^{(\Phi_x,\Phi_y)}}{\mathcal{Z}},
\end{equation}
where we have defined a sub-partition function for each global
flux sector $(\Phi_x,\Phi_y).$ So, for the uniform flux states
$\mathcal{Z}^{(\Phi_x,\Phi_y)} = {\rm tr}\big|_{(\Phi_x,\Phi_y)}\
\exp(-\mathcal H/T)$.

Clearly, in the absence of a dependence of the {\it physical}
spectrum on the global flux, all the sub-partition functions will
be identical $\mathcal{Z}^{(-1,-1)}=
\mathcal{Z}^{(1,1)}=\mathcal{Z}^{(1,-1)}=\mathcal{Z}^{(-1,1)}$ and
$\langle \Phi_\alpha \rangle$ will average out to be identically
zero. This is the case for the $A$ phase; and the case of toric
code previously studied\cite{fragility}. However, for the $nA$
phase of the CSL model, the $(-1,-1)$ sector is projected out of
the ground state Hilbert space and hence
$\mathcal{Z}^{(-1,-1)}(T=0)_{\rm nA}=0$. This yields a {\it
finite} and definite $\langle \Phi_\alpha \rangle$ in the $nA$ phase
at $T=0$:
\begin{equation}
\langle \Phi_x \rangle({T=0}) =\left\{
\begin{split}
1/3\quad ({\rm nA}, g<g_c)\\
0\quad ({\rm A}, g>g_c)
\end{split}\right. .
\label{EQ:NAExpect}
\end{equation}
The significance of Eq.\eqref{EQ:NAExpect} is that
 $\langle \Phi_x \rangle$ is the first identification of a quantity that can be defined in a thermodynamic sense that
 {\it changes} at the topological QPT.
Now the relation $\langle \Phi_x \rangle(T=0)$ can be related to
the topological degeneracy  through
\begin{equation}
n_{DEG} = 4 - 3\langle \Phi_x \rangle(T=0).
\label{EQ:thermalDeg}
\end{equation}
Most importantly,
 this identification allows one to extend  the notion of topological degeneracy to
 {\it finite temperature} through $\langle \Phi_x \rangle(T\neq0)$ and to investigate
 the vicinity of the topological QPT that is largely unknown in Fig.1(b).

\begin{figure}
\centerline {
\includegraphics[width=.35\textwidth]{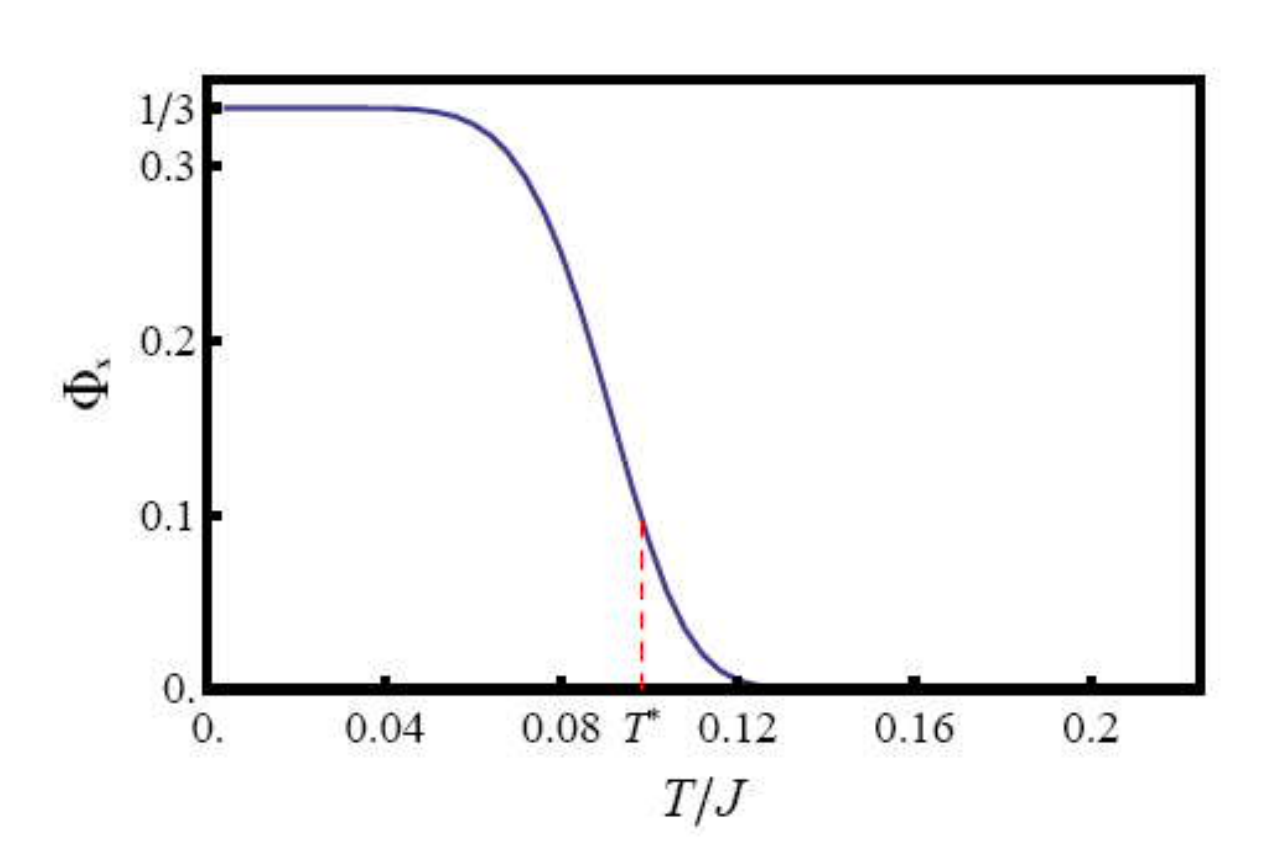}
}
\caption{Defining $T^*$  through the exponential decay of the $\langle \Phi_x (T)\rangle$. The plot is for $g=1.3< g_c$ on a $60\times 40$ lattice.} 
\label{FIG:crossoverSharp}
\end{figure}

\begin{figure}
\centerline {
\includegraphics[width=.35\textwidth]{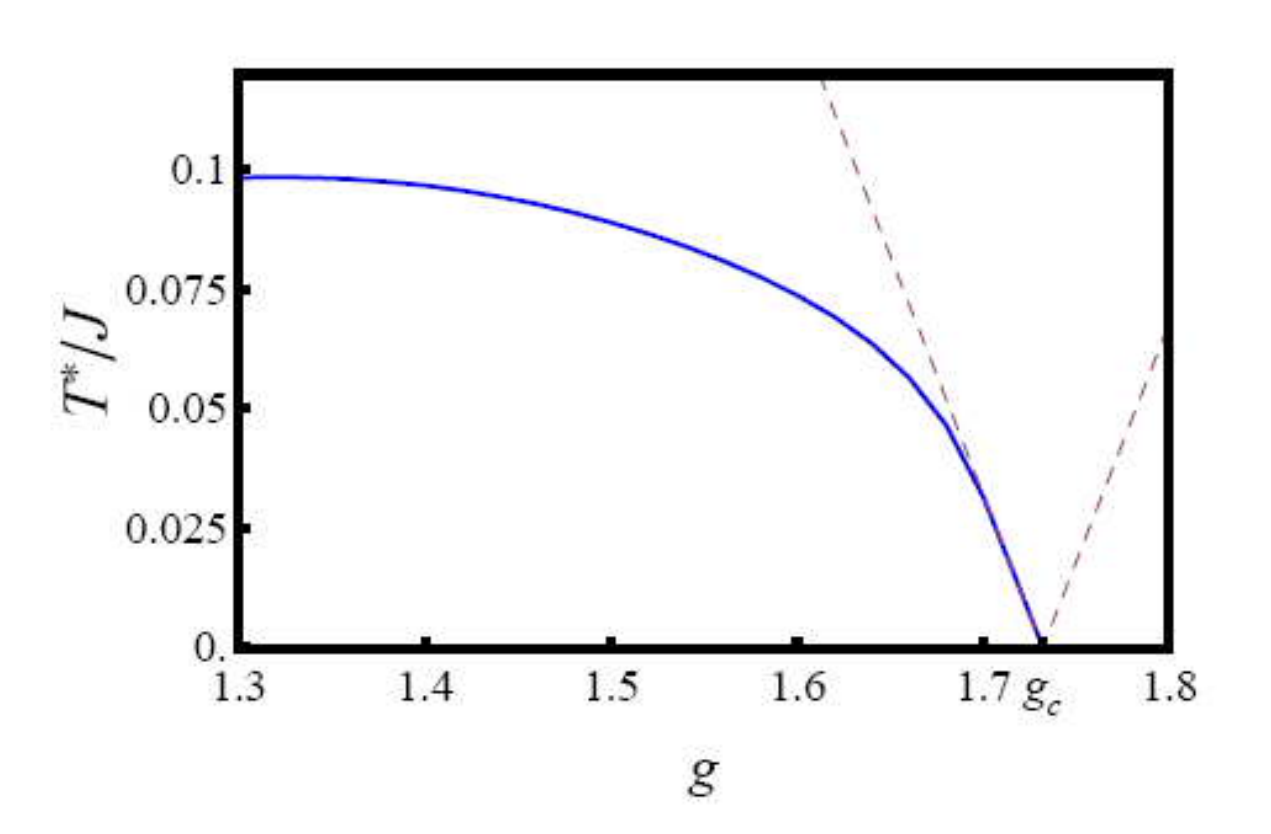}
}
\caption{The  $T^*$(solid line) compared to the single fermion excitation gap (dashed line) for a 60$\times$40 lattice in the vicinity of $g_c $. Although both Abelian and non-Abelian phases are gapped, $T^*$ is defined only in the non-Abelian phase.
}
\label{FIG:crossover}
\end{figure}

Analogous to $T_c(g)$ (the solid line of Fig.1(a)) and the cross
over line (dashed line in Fig.1(a)) in the vicinity  of a
conventional QCP, we define and investigate a cross over
temperature scale $T^*(g)$ above which $\langle \Phi_x \rangle$
falls off to zero 
for a system with finite size. As shown in Fig.~(\ref{FIG:crossoverSharp}), a
crossover temperature scale $T^*(g)$ can be defined as the
temperature scale at which
 $\langle \Phi_x \rangle(T)$ falls off exponentially from its
 zero temperature value at a given value of $g<g_c$.
 In Fig.~(\ref{FIG:crossoverSharp}), we show
 $\langle \Phi_x \rangle(T)$ for $g= 1.3$ as defined in Eq.\eqref{eq:Phi}.
  For all $g<g_c$, $\langle \Phi_x \rangle$ is nearly a constant $(1/3)$ for $T<T^*(g)$, but decays exponentially to zero at higher
  temperatures.
  We have defined $T^*(g)$  as the point at which $\mathcal{Z}^{(-1,-1)}(T^*)/\mathcal{Z}^{(1,1)} (T^*)=e^{-1}$ by convention.
  Above $T^*$, the distinction between the $A$ and the $nA$ phase vanishes. The plot of $T^*(g)$ in Fig.\ref{FIG:crossover} shows that  $T^*(g)$ is a
  distinct scale which is non-vanishing at the QCP $g=g_c,$ unlike the excitation gap which
  vanishes.  The excitation gap is non-vanishing in both phases where, in contrast, $T^*(g)$ is only non-zero for $g<g_c$ which allows the identification of the $nA$ phase at finite temperature.

The crossover scale $T^*$ has an intriguing system size dependence. For a large system of $N$ sites,
\begin{equation}
T^* \sim \frac{\Delta(g)}{\ln N},
\label{EQ:crossover}
\end{equation}
where  $\Delta(g)$ is the energy gap in the fermion spectrum(see
Fig.\ref{FIG:crossover}). Hence $T^*\to 0$ as  $N\to\infty$, but
at a rate slower than any other quantity in the system. Therefore,
the distinction between the A phase and the $nA$ phase, while
strictly vanishing at finite temperature in the thermodynamic
limit, can be meaningful in some sizable range of $N$.
Interestingly, this system size dependence bares similarity with
the crossover scale for the finite temperature topological
entanglement entropy of the toric code in
Ref.\cite{thermal_Claudio}. 
It is also reminiscent of the finite temperature behavior of 
an Ising chain of finite length. An  Ising chain does not order at 
any finite temperature in the thermodynamic limit. However, one 
can define a finite crossover temperature scale in a finite size 
system of length $N$  by comparing the energy cost $2J$ for a 
domain wall and the entropic gain of $T\log N$ for $N$ possible 
choice for the position of the domain wall. Nevertheless, the present  
size dependence Eq.\eqref{EQ:crossover} is rather a consequence 
of $\langle \Phi_x \rangle(T)$ being bounded from below by 
$\tanh^N(\Delta/2T)/3$ (see SOM) and it is unrelated to local fluctuations.
Possible connections underlying these apparent similarities and their 
implications for a ``quantum critical region'' is an open question.


\begin{figure}
\centerline {
\includegraphics[width=.35\textwidth]{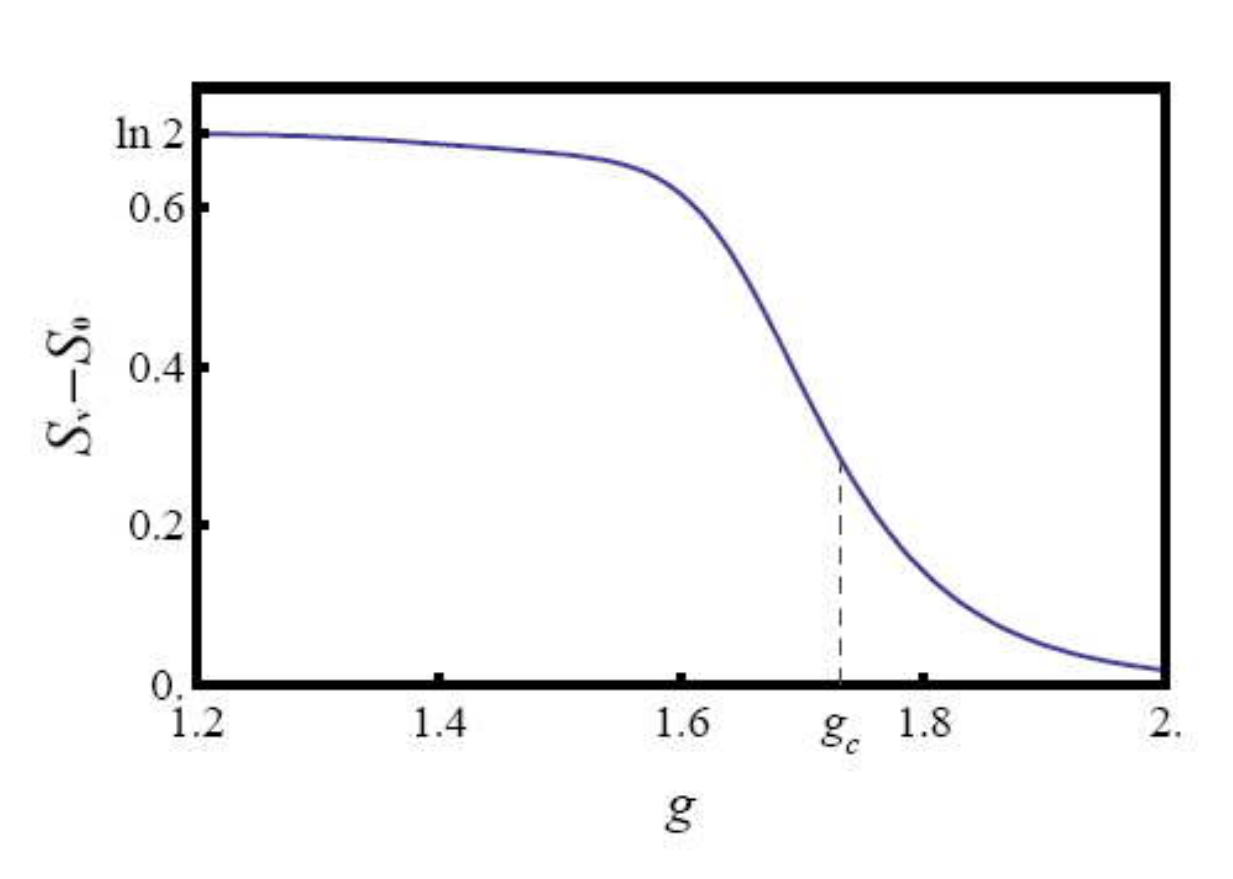}
}
\caption{The entanglement entropy change due to a vortex pair excitation for 30 $\times$ 30 Kagom$\acute{e}$ lattice sites on a torus. The change at $g_c$ is not sharp 
 for this finite size, 
though it approaches the thermodynamic limit value of $\ln 2$ in the $nA$ phase.}
\label{FIG:entangle}
\end{figure}

{\bf Towards finite $T$ entanglement entropy}--While the
topological entanglement entropy \cite{kitaev-preskill,levin-wen}
of the ground state wave function has been widely adopted as a
measure of topological order, it is does not distinguish the A
phase from the $nA$ phase in the present case \cite{zhenghan}. The topological
entanglement entropy $\gamma$ is defined as the universal constant
term in the entanglement entropy in addition to the usual term
proportional to the perimeter of the boundary: $S_{\rm ent}=\alpha
L -\gamma$. Further it is known that $\gamma$ is given by the
total quantum dimension of the topological phase :
$\gamma=\ln\sqrt{\sum_\alpha d_\alpha^2}$. Since $\{d_\alpha\} =
\{1,\sqrt{2},1\}$ in the $nA$ phase while $\{d_\alpha\} =
\{1,1,1,1\}$ in the $A$ phase \cite{zhenghan}, the topological QPT does not
affect $\gamma$ which is equal to $2$ in both phases. However, it
is possible a finite temperature extension of this quantity might
offer a possible distinction between the two phases, as
excitations with distinct statistics would contribute.

An extension of the entanglement entropy to $T\neq0$ must involve
the inclusion of thermal excitations. For instance, the extension
proposed in Ref.\cite{thermal_Claudio} (an alternative definition 
was proposed in Ref.\cite{pachos}) retains the basic form 
$S_{\rm ent}=-{\rm Tr}(-\rho_{\mathcal{A}} \ln \rho_{\mathcal{A}})$ 
but uses for the $\rho_{\mathcal{A}}$, a thermalized reduced 
density matrix
\begin{equation}
\rho_{\mathcal{A}}(T) = \sum_{\lambda} \frac{e^{-E_\lambda/T}}{Z} {\rm Tr}_{\mathcal{B}}
|\Phi_\lambda\rangle \langle \Phi_\lambda |,
\label{eq:SeT}
\end{equation}
where $|\Phi_\lambda\rangle$'s are energy eigenstates. Clearly
Eq.\eqref{eq:SeT} reduces to the usual definition at $T=0$ when
only the ground state(s) enter the sum.

While the zero temperature topological entanglement entropies of the $A$ and $nA$
phases are identical, the different excitations of each phase will
generically lead to different finite-temperature quantities. As a
first pass through the problem, we consider the change in the
entanglement entropy in the presence of a pair of vortex
excitations, one each in the two regions $\mathcal{A}$ and
$\mathcal{B}$ whose entanglements are under consideration.
Fig.~(\ref{FIG:entangle}) shows the result as a function of $g$.
The additional  entropy of $\log 2$ of the $nA$ phase reflects the
double degeneracy associated with  Majorana fermion vortex core
states responsible for the non-Abelian statistics of the vortices.
It is clear that the characteristics of finite-energy excitations
are important qualities of the topological phases.

{\bf Closing remarks}-- We studied the nature of the topological
QPT between a $nA$ phase and an $A$ phase in an exactly solvable
model using finite temperature extensions of two separate measures
of topological order.   The expectation value of the global flux
$\langle \Phi\rangle(T)$  is a finite temperature extension of the
ground state topological degeneracy which clearly changes at
the QPT. We found that  $\langle \Phi\rangle(T)$ retains the $T=0$
value for $g<g_c$ up to a crossover temperature scale which decays
logarithmically with the system size. Whether this type of
crossover is ubiquitous for topological phases in two spatial
dimension is an open question.  As a step towards a finite
temperature extension of $\gamma,$ which is independent of $g$ at
$T=0$, we considered the effect of a pair of vortices. This
indicates the possibility that $\gamma(T)$ might distinguish the
$nA$ phase from the $A$ phase.

{\bf Acknowledgement}- We would like to thank C. Castelnovo, 
C. Chamon,  S. Kivelson, Z. Nussinov, G. Ortiz, M. Oshikawa, 
J. Pachos, S. Sondhi, M. Stone, S. Trebst, Z. Wang, T. Xiang 
for sharing their insights. 
SBC and TLH are supported by DOE grant DE-AC03-76SF00515; SBC also by SITP at Stanford University. HY is supported by DOE grant DEFG03-01ER45925 and SGF at Stanford University. E-AK is supported in part by the Nanoscale Science and Engineering Initiative of the National Science under NSF Award \#EEC-0646547.
We acknowledge the UIUC ICMT for its hospitality through 
``Workshop on Topological Phases in Condensed Matter" during 
early stages of this work.


\appendix
\section{The supporting online material: The system size dependence of $T^*$ in the non-Abelian phase.}
Since the vortex gap never closes for all values of $g$,  we can ignore the vortex excitation in calculating $\langle \Phi_x \rangle$ in the vicinity of the QPT:
\begin{equation}
\langle \Phi_x \rangle(T) = \frac{{\cal Z}_{CSL}^{(1,1)} - {\cal Z}_{CSL}^{(-1,-1)}}{3{\cal Z}_{CSL}^{(1,1)} + {\cal Z}_{CSL}^{(-1,-1)}}. 
\label{eq:Phi-uf-Z}
\end{equation}
Since in the $nA$ phase, the fermion occupation number parity
 is odd in the $(\Phi_x,\Phi_y) = (-1,-1)$ sector and even in all the other sectors
\begin{equation}
{\cal Z}_{CSL}^{(1,1)} \pm {\cal Z}_{CSL}^{(-1,-1)}= 2 \exp(- E_G/T) \prod_{n, \vec{k}} [1 \pm \exp(-\epsilon_{n,\vec k}/T)],
\label{eq:even-odd}
\end{equation}
where $E_G$ is the ground state energy. With Eq.\eqref{eq:even-odd}, we can recast the global flux thermal expectation value $\langle \Phi_x \rangle(T) $   as
\begin{equation}
\langle \Phi_x \rangle(T) = \frac{\prod_{n, \vec{k}} \tanh( \epsilon_{n,\vec k}/2T)}{2 + \prod_{n, \vec{k}} \tanh(\epsilon_{n,\vec k}/2T)}.
\end{equation}
This implies
\begin{equation}
\frac{1}{3}\left[\tanh\frac{\Delta(g)}{2T}\right]^N < \langle \Phi_x \rangle < \frac{1}{2}\left[\tanh\frac{\Delta(g) + W(g)}{2T}\right]^N
\label{EQ:pigeonExpect}
\end{equation}
where $\Delta(g)$ and $\Delta(g)+W(g)$ are respectively the minimum and maximum of the energy spectrum$\epsilon_{n, \vec{k}}[g]$ that depends on g. 
Since $\Delta(g)$ and $W(g)$ remains finite in the thermodynamic limit of $N \to \infty$ 
Eq.\eqref{EQ:pigeonExpect} implies 
\begin{equation}
\lim_{N\to \infty}\langle \Phi_x\rangle \to 0
\end{equation}
in the thermodynamic limit.
Further using $[\tanh (\Delta/2T)]^N \approx 1-2N\exp(-\Delta/T)$ at large $N$ and low $T$, we find 
the system size dependence of the crossover temperature:
\begin{equation}
T^* \sim \frac{\Delta(g)}{\ln N},
\end{equation}

\end{document}